\documentclass{revtex4}
\usepackage{amsmath}
\usepackage{epsfig}
\newcommand{\adu}{d^\dagger_{i\uparrow}}
\newcommand{\alu}{d^{\vphantom{\dagger}}_{i\uparrow}}
\newcommand{\px}{p^{\vphantom{\dagger}}_{i+\hat{x}/2,\sigma}}
\newcommand{\py}{p^{\vphantom{\dagger}}_{i+\hat{y}/2,\sigma}}
\newcommand{\pmx}{p^{\vphantom{\dagger}}_{i-\hat{x}/2,\sigma}}
\newcommand{\pmy}{p^{\vphantom{\dagger}}_{i-\hat{y}/2,\sigma}}
\newcommand{\Pl}{P^{\vphantom{\dagger}}_{i\sigma}}
\newcommand{\Pdu}{P^\dagger_{i\uparrow}}
\newcommand{\Plu}{P^{\vphantom{\dagger}}_{i\uparrow}}
\newcommand{\up}{\uparrow}
\newcommand{\dw}{\downarrow}
\newcommand{\tr}{{\mathrm{tr}}\;} 

\begin{document}


\title{Fermion kinetics in the Falicov-Kimball limit of the three-band Emery
model}

\author{D. K. Sunko}
\affiliation{Department of Physics, Faculty of Science, Bijeni{\v c}ka 32,
HR-10000 Zagreb, Croatia.}

\begin{abstract}
The three-band Emery model is reduced to a single-particle quantum model of
Falicov-Kimball type, by allowing only up-spins to hop, and forbidding double
occupation by projection. It is used to study the effects of geometric
obstruction on mobile fermions in thermodynamic equilibrium. For low hopping
overlap, there appears a plateau in the entropy, due to charge correlations,
and related to real-space disorder. For large overlap, the equilibrium
thermopower susceptibility remains anomalous, with a sign opposite to the one
predicted from the single-particle density of states. The heat capacity and
non-Fermi liquid response are discussed in the context of similar results in
the literature. All results are obtained by evaluation of an effective
single-particle free-energy operator in closed form. The method to obtain this
operator is described in detail.
\end{abstract}

\maketitle

\section{Introduction}

There is at present no paradigmatic `non-Fermi liquid' in more than one
dimension, in the same sense as there is the Luttinger liquid in the 1D case.
It is therefore of continuous interest to study failures of the Fermi liquid
concept in special cases, accessible to concrete calculation. The behavior of
electrons in the presence of localized obstructions has long been a fertile
ground for investigating deviations from simple Fermi liquid behavior.
Examples are the quantum Hall effect~\cite{Klitzing86}, the Kondo
problem~\cite{Andrei83}, localization by impurities~\cite{Anderson78}, and the
Mott metal-to-insulator transition~\cite{Mott68}. All of these save the last
involve fixed impurity distributions, and their unusual behavior is related to
the creation of localized states in the presence of scatterers. Because of
this, states accessible to weak probes cannot be understood in the picture of
nearly independent quasiparticles.

The Mott transition is unique among the above because the localized
obstructions are the mobile electrons themselves, due to the strong on-site
repulsion between electrons of the opposite spin. This means translational
invariance is not broken, and obstruction is dynamical, which makes the
problem much more difficult than with quenched disorder. Despite many
contributions which have shaped our present
understanding~\cite{Hubbard63,Gutzwiller65,Brinkman70,Anderson78-1,
Kotliar86,Kotliar88,Metzner89,Georges96}, no picture of the Mott transition
has emerged to date, which convincingly describes the phenomenon in space and
time.

Motivated by the Mott problem in the charge-transfer limit, which is of
primary interest for high-temperature superconductors~\cite{Anderson87}, the
present work describes non-Fermi liquid behavior in one model of the
Falicov-Kimball~\cite{Falicov69} group, which are intermediate between the
quenched-disorder case and the fully dynamical Mott case: the disorder is
heavy, but annealed. In this way translational invariance is restored at the
level of the ensemble, while the dynamics still relates to static disorder.
The present model is the only one in the group in which the disorder is truly
geometric, \emph{i.e.} quantum processes do not physically interact with any
classical degree of freedom. This makes it an interesting test-bed for
non-Fermi liquid behavior, and indeed it is found that geometric obstruction
induces collective behavior in the `strange metal' state beyond half-filling.
It somewhat resembles normal $^3$He at 0.5~K, where the entropy saturates at
$\ln 2$ per atom~\cite{Seiler86}, simply because the atoms push against each
other, obstructing kinetic motion. In the limit of low hopping overlap, the
model develops an unexpected similarity to the Kauzmann
paradox~\cite{Kauzmann48} in vitreous liquids, despite the fact that all
calculations are at equilibrium, so one cannot speak of kinetic slow-down in
the usual sense.

For intermediate-to-large overlaps, the model appears at first sight as a
renormalized Fermi liquid. However, the `equilibrium thermopower,'
$\partial\mu/\partial T$, behaves exactly oppositely to what is predicted from
the effective one-particle density of states (1$p$-DOS) at the same
temperature and filling. This persistence of quantum collectivity is observed
because geometric obstructions take the form of boundary conditions, so they
cannot be suppressed by hopping fluctuations.

The technique used to analyze the equilibrium properties of the model is
presented here for the first time. It enables one to find a closed analytical
expression for the free energy. The essential idea is to remain in the Fock
space of the mobile fermions, while evaluating the logarithm of the annealed
partition function. The one-particle part of this logarithm is then itself a
one-particle operator in Fock space, and can be diagonalized exactly, just
like any tight-binding Hamiltonian. With such a careful treatment of Pauli
correlations, it is sufficient to expand the annealed free energy to the first
non-trivial term in a Kubo cluster cumulant expansion~\cite{Kubo62}. The
resulting analytical expression is exactly correct in both the atomic and
metallic limits, and qualitatively compares well with numerical calculations
on similar models.

\section{The model}

\subsection{Brief description}

\begin{figure}[tb]
\center{\epsfig{file=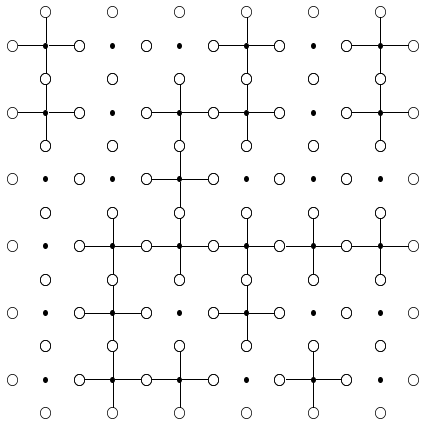,height=55mm}}
\caption{The random-tiling model. Black lines are hopping integrals for
up-spins, connecting copper (black) and oxygen sites (white). The hopping
integrals are turned off where there is a down-spin on a copper site.}
\label{fig_RTM}
\end{figure}

The geometric Falicov-Kimball model used here has been introduced
previously~\cite{Sunko96}. It is given by the following modification of the
well-known Emery (three-band) model~\cite{Emery87}:
\begin{equation}
H=\varepsilon_d\sum_{i,\sigma}\hat{n}_{i\sigma}+
\varepsilon_p\sum_{i,\sigma}(\hat{n}_{x,i\sigma}+\hat{n}_{y,i\sigma})+
t\sum_{i}(1-\hat{n}_{i\dw})\left(\adu\Plu+\Pdu\alu\right)\,,
\label{rt}
\end{equation}
where $\Pl=\pmx+\pmy-\px-\py$ describes the four oxygen $p$-orbitals around a
copper $d$-orbital, and $\hat{n}_{i\sigma}$, $\hat{n}_{x,i\sigma}$,
$\hat{n}_{y,i\sigma}$ are the number operators of the copper and oxygen sites.
Notice that only up-spins can hop. This is a simple realization of the idea
behind Gutzwiller's approximation: fermions of one spin orientation see those
of the other `as if occupying a band of width zero'~\cite{Gutzwiller65}. The
static down-spins are scattered at random. If one lands on an oxygen site, it
pays the price of a charge-transfer energy
$\Delta_{pd}=\varepsilon_p-\varepsilon_d>0$ between oxygen and copper, but is
otherwise inactive. Thus the mobile up-spins see a lattice with defects,
different for each arrangement of down-spins. The situation is summarized in
Figure~\ref{fig_RTM}. (For those familiar with the Emery model, note that the
up- and down-spins here correspond to physical holes in the copper
$d$-orbitals, not electrons.)

In this variant of the Falicov-Kimball~\cite{Falicov69} model, there is no
physical scattering off the heavy classical spins. Instead the mobile spins
see a `torn' lattice with open boundary conditions where the hopping has been
turned off. Thus they are in the strong-coupling limit on the one hand, and
there is no singularity in their Hamiltonian, on the other.  Now there
intervenes a significant topological simplification, namely the `crosses' of
hopping overlaps impinging on copper sites tile the plane, as visible in
Fig.~\ref{rt}. This gives the model its name, random-tiling (RT) model, and
makes it possible to express the annealed free energy of the problem by a Kubo
cluster-cumulant expansion without leaving the Fock space of
up-spins~\cite{Sunko96,Sunko00-1}. Double occupation of copper sites is
disallowed by a separate factor $\exp(-\beta U \sum_i
\hat{n}_{i\up}\hat{n}_{i\dw})$ in the partition function, with $U\gg
\Delta_{pd}$. This factor is not entangled with the Hamiltonian, so it is not
an interaction, but pure bookkeeping, to force all the mobile spins away from
the `torn' part of the lattice. The projector $(1-\hat{n}_{i\dw})$ in
Eq.~(\ref{rt}) is treated exactly, which is essential~\cite{Millis91}: both
the Pauli principle and the geometric obstruction are treated on an equal
footing. One projects the Fock-space Kubo expansion onto the single-particle
subspace, and keeps the first non-trivial term, while higher single-particle
terms can be shown to be negligible. This gives a one-particle quantum
free-energy operator, whose spectrum can be exactly obtained in closed form.
It is built from up-spin wave functions which anneal the geometric disorder.
It provides the effective background for all residual (two-particle)
interactions, so it is of interest to study by itself. The Mott-Hubbard
transition has already been studied in some detail~\cite{Sunko00}.

\subsection{Detailed derivation}

The present section is purely technical, and quite independent of the rest of
the article. It describes how the annealed free energy of the random-tiling
model can be written without leaving the Fock space of up-spins. The
one-particle part of this expression takes the form of a rapidly convergent
cluster cumulant expansion around the atomic limit. The main strength of the
procedure is that it never violates either the geometric or the Pauli
constraints. In particular, it is shown that the same closed expression
exactly recovers the metallic free-fermion limit for the up-spins, when the
down-spin concentration goes to zero. This is the \emph{opposite} limit from
the atomic one, which attests to the advantage of working in Fock space. Such
physical considerations are collected in the last part of this section, to
which one may prefer to skip at first reading.

\subsubsection{Problem and method}

The physical picture described above corresponds at equilibrium to the
evaluation of the canonical partition function
\begin{equation}
Z=\tr e^{-\beta H}
\end{equation}
where $H$ is the Hamiltonian~(\ref{rt}), under the geometric constraint, that
mobile up-spins do not land on the copper sites occupied by the down-spins, to
which the hopping has been turned off by the projector in~(\ref{rt}). As
already mentioned above, this is implemented by actually evaluating the
unconstrained trace of
\begin{equation}
Z=\tr e^{-\beta H}e^{-\beta V}=
\tr e^{-\beta V/2}e^{-\beta H}e^{-\beta V/2},
\label{UZ}
\end{equation}
where the disentangled term $e^{-\beta V}$ enforces the constraint, with
\begin{equation}
V=U\sum_{i} \hat{n}_{i\up}\hat{n}_{i\dw}.
\end{equation}%
(The form with $V/2$ turns out to be more convenient for later manipulation.)
It puts the mobile up-spins onto that part of the lattice which is not `torn'
by the down-spins on copper sites, acting through the projector
$1-\hat{n}_{i\dw}$ in Eq.~(\ref{rt}). One can easily put $U\to\infty$ in the
final expressions, but it is neater to keep a finite $U\gg\Delta_{pd}$. As
expected, the results are numerically independent of $U$ in this limit.

The trace over down-spins in Eq.~\ref{UZ} is trivial to perform, and results
in rather laborious expressions of the generic form
\begin{equation}
\sum_{\{i_1,i_2,\ldots,i_M\}\subset{\mathcal L}}
f(A_{i_1}+\ldots +A_{i_M})\equiv U_{LM}.
\label{ulm}
\end{equation}
Here $A_i$ are site-labelled operators in the Fock space of up-spins,
essentially the crosses in Fig.~\ref{fig_RTM}. This is obvious, since a given
configuration of down-spins corresponds to some choice of $M$ lattice sites
where the hopping survives, out of a total $L$ sites ($L$ is volume)
comprising the lattice ${\mathcal L}$. Note that $f$ is a function of
Fock-space operators for the up-spins. It is something like
\begin{equation}
f(X)\sim\exp\left[-\beta(H_0+X)\right],
\end{equation}
where $H_0$ is the site Hamiltonian. The $U_{LM}$ are the thermal evolution
operators for the annealed disorder with $M$ hopping crosses present. The main
point of the method used here is to reorganize expressions like (\ref{ulm}) by
a combinatorial inversion which is valid for arbitrary non-commuting
operators. This allows one to extract the one-particle part of the free energy
expansion exactly, and show that corrections to its first (one-site) term are
controlled in the range of fillings of interest here.

Physically, the method boils down to a combinatorial interpretation of Kubo's
cluster cumulant formula for the free energy. This is explained at length in
Ref.~\cite{Sunko00-1}, where these ideas are used to cover familiar ground. In
particular, the reader is reminded that the expansion parameter, corresponding
to the concentration in the classical virial expansion, is an occupation
probability in the general case~\cite{Kubo62}. In the present work, one
implements Kubo's formalism by first performing the combinatorial inversion at
the level of the stochastic evolution operators~(\ref{ulm}), and taking the
logarithms only in the last step. To be explicit, introduce the binomial
transform
\begin{subequations}
\begin{eqnarray}
U_{LM}&=&\sum_p \binom{L-p}{M-p} U_L(p),
\label{invb}
\\ U_L(p)&=&\sum_M (-1)^{p-M}\binom{L-M}{p-M} U_{LM}.
\end{eqnarray}
\end{subequations}
When one of these expressions is inserted into the other, the result is an
identity, so they are valid independently of what the $U_{LM}$ may be. Then a
very short exercise shows that the generating function for the $U_{LM}$'s may
be expressed in terms of the $U_L(p)$ as
\begin{equation}
\sum_M x^M U_{LM} = (1+x)^L\sum_p y^p U_L(p),
\label{fiz}
\end{equation}
where $y=x/(1+x)$, $0<y<1$. This result depends only on the relation
(\ref{invb}), and in particular, nothing is assumed about commutativity. Since
the $U_{LM}$ are canonical evolution operators, the sum in (\ref{fiz}) is the
corresponding grand canonical operator. Hence $x$ involves the chemical
potential for the hopping crosses, $x=\exp(\beta\nu')$, say. In the present
context, it will be more convenient to express the expansion parameter $y$ in
terms of the chemical potential of the static down-spins, call it $\nu$. Since
the hopping cross is absent when the down spin is present on a copper site,
the two chemical potentials are effectively connected by a particle-hole
transformation. When the energy origin is conveniently chosen so that
$\varepsilon_d=0$, this amounts to $\nu=-\nu'$, i.e.
\begin{equation}
y=\frac{1}{1+e^{\beta\nu}}.
\end{equation}
For the range of concentrations considered in this article, half-filling and
beyond, the chemical potential of the static down-spins will turn out to be at
the oxygen energy, $\nu\approx\varepsilon_p>0$, so that $y$ is exponentially
small. Thus it provides rapid convergence in the expansion of the one-particle
part of Eq.~(\ref{UZ}) for these concentrations.

\subsubsection{Main expressions}

Set the bare copper energy $\varepsilon_d=0$. In Emery-model jargon, the `hole
picture' is assumed, so $\varepsilon_p>\varepsilon_d=0$, and as mentioned
above, $U\gg \varepsilon_p$. Call
\begin{equation}
\widehat{N}_{d\uparrow}=\sum_{i=1}^Ln_{i\uparrow}
\end{equation}
the number operator on the copper sites, which appears in (\ref{rt}). The
actual forms encountered in tracing (\ref{UZ}) are slightly more complicated
than suggested in the previous section, because of the side factors with $V/2$,
and because the down spins can also land on the oxygen sites. Overcoming
these complications by ordinary diligence, the $U_L(p)$, required in
(\ref{fiz}), are given by
\begin{eqnarray}
U_L(0)&=&e^{-\beta\left(\varepsilon_p\widehat{N}_{p\uparrow}
+U\widehat{N}_{d\uparrow}
\right)},
\label{nulti}\\ U_L(1)&=&\sum_{i=1}^L\left( e^{-\beta
(U/2)(\widehat{N}_{d\uparrow}-n_{i\uparrow})}
e^{-\beta[\varepsilon_p\widehat{N}_{p\uparrow}+t(
d^{\dagger}_{i\uparrow}P^{\vphantom{\dagger}}_{i\uparrow}
+P^{\dagger}_{i\uparrow}
d^{\vphantom{\dagger}}_{i\uparrow})]}
e^{-\beta (U/2)(\widehat{N}_{d\uparrow}-n_{i\uparrow})}
-U_L(0)\right).
\label{prvi}
\end{eqnarray}
Here, $\widehat{N}_{p\uparrow}$ is the number operator on the oxygen sites.
Stopping at first order in $y$ in the general expansion (\ref{fiz}), the
operator to be traced over the up-spins now appears as
\begin{eqnarray}
\lefteqn{
\sum_{M_\downarrow}\binom{qL}{N_\downarrow-M_\downarrow}
e^{-\beta\varepsilon_p(N_\downarrow-M_\downarrow)}
\nonumber}\\
&&\times\left[\left[x^{L-M_\downarrow}\right]\right]
(1+x)^L\exp\left\{\left[\ln\left(U_L(0)+yU_L(1)\right)\right]_{1p}\right\},
\label{op}
\end{eqnarray}
where the notation `$[[x^r]]\ldots$' means `coefficient of $x^r$ in~$\ldots$',
and `$1p$' means `one-particle part.' Here $N_\downarrow$ is the total number
of static down-spins, and $M_\downarrow$ is the number of down-spins which
have landed on the coppers. The number of oxygen sites per unit cell is $q=2$.
In the next section, the logarithm of Eq.~(\ref{op}) will be explicitly
computed in Fock space. Assume for now that this has been done, and that the
one-particle operator in braces in (\ref{op}) has been diagonalized. 
Then the trace can also be performed in the space of up-spins. Introducing a
chemical potential $\mu$ for these latter, the trace now reads
\begin{equation}
\sum_{M_\downarrow}\binom{qL}{N_\downarrow-M_\downarrow}
e^{-\beta\varepsilon_p(N_\downarrow-M_\downarrow)}
\frac{(1+e^{-\beta\nu})^LZ(\mu,\nu)}
{e^{\beta\mu N_\uparrow}e^{-\beta\nu(L-M_\downarrow)}},
\label{trop}
\end{equation}
where $x$ has been called $e^{-\beta\nu}$, $\nu$ is the chemical potential of
the down-spins, and
\begin{equation}
Z(\mu,\nu)=\prod_{k,\alpha}\left(1+e^{\beta(\mu-\varepsilon_\alpha(k,\nu))}
\right)
\label{zmunu}
\end{equation}
is the grand partition function in up-spin space, with
$\varepsilon_\alpha(k,\nu)$ the energies of the up-spin normal modes, indexed
with a band index $\alpha$, and depending on the chemical potential of the
down-spins, $\nu$, as well as on a wave vector, $k$. Finally, the sum in
(\ref{trop}) is evaluated by taking its largest term, resulting in the
following model: find the extremum of
\begin{equation}
\sqrt[L]{Z_{\mathrm{eff}}(\mu,\nu)}=\frac{(1+e^{\beta\nu})
(1+e^{\beta(\nu-\varepsilon_p)})^q\sqrt[L]{Z(\mu,\nu)}}
{e^{\beta\mu n_\uparrow}e^{\beta\nu n_\downarrow}},
\label{model}
\end{equation}
in the space of $\mu$ and $\nu$, for given concentrations $n_\uparrow$ and
$n_\downarrow$. The extremal equations are
\begin{subequations}
\label{eqmunu}
\begin{eqnarray}
n_\uparrow&=&\frac{1}{L}\frac{\partial\ln Z(\mu,\nu)}{\beta\partial\mu},
\\ n_\downarrow&=&\frac{e^{\beta\nu}}{1+e^{\beta\nu}}
+q\frac{e^{\beta(\nu-\varepsilon_p)}}{1+e^{\beta(\nu-\varepsilon_p)}}
+\frac{1}{L}\frac{\partial\ln Z(\mu,\nu)}{\beta\partial\nu}.
\label{eqnu}
\end{eqnarray}
\end{subequations}
Eq.~(\ref{eqnu}) is physically transparent: there are static down-spins,
sitting in one level at $\varepsilon_d=0$, and $q=2$ levels at the oxygen
energy $\varepsilon_p$. The last term, affecting the count of down-spins,
comes from the interaction with the up-spins, as coded by the dependence of
their dispersion on $\nu$. This dispersion is derived in the next section,
giving a complete, closed model. Its dependence on the chemical potential of
the down-spins is expected on general thermodynamic grounds, since the
presence of down-spins affects the configurations available to up-spins.

\subsubsection{Extraction of the one-particle part}

In the previous section, the binomial transform was used to treat the
stochastic evolution operator induced by the geometric constraint
(annealment). The manipulations were mostly an exercise in bookkeeping. In
this section, the Pauli constraint is considered, and the main tool is the
evaluation of analytic functions of matrix arguments. This is the
straightforward way to evaluate analytic functions of Fock space operators, if
one is only interested in the resulting one-particle part. Analytic functions
of matrix arguments are evaluated as usual, by diagonalization:
$$
F(A)=F(SS^{-1}ASS^{-1})=SF(S^{-1}AS)S^{-1}.
$$
To prepare the ground requires, however, one last combinatorial trick. Namely,
take the $f(X)$ in (\ref{ulm}) to be $\ln(U_L(0)+yX)$. Then
$\ln(U_L(0)+yU_L(1))$ can play the role of an `$U_{LL}$' in (\ref{ulm}), since
$U_L(1)$ in (\ref{prvi}) is a sum over \emph{all} site-indexed terms, each
playing the role of an `$A_i$' in (\ref{ulm}). The expansion (\ref{invb}) with
$M=L$ then gives, for this case,
\begin{eqnarray}
\lefteqn{
\ln\left(U_L(0)+y\sum_{i=1}^LB_i\right)=
-\beta(\varepsilon_p\widehat{N}_{p\uparrow}+U\widehat{N}_{d\uparrow})
}\nonumber\\
&&+\sum_{i=1}^L\left\{\ln\left[U_L(0)+yB_i\right]-\ln\left[U_L(0)\right]
\right\}
+\sum_{i<j}^L\ldots,
\label{ln}
\end{eqnarray}
where the sum over $i<j$ is already a second-order term. Here $B_i$ denotes
the $i$-th term under the sum in (\ref{prvi}). The effective free energy is
then
\begin{equation}
F_{\mathrm{eff}}=\varepsilon_p\widehat{N}_{p\uparrow}+U\widehat{N}_{d\uparrow}
-\frac{1}{\beta}\sum_{i=1}^L\left\{\ldots\right\}_{1p},
\label{heff}
\end{equation}
where the braces refer to the same expression as in Eq.~(\ref{ln}). Higher
order terms are dropped, because they correspond to higher powers in $y$.

From now on the way is clear, provided one has an algebraic manipulation
program. The effective free energy (\ref{heff}) may be rewritten
\begin{eqnarray}
F_{\mathrm{eff}}&=&\sum_i\left.\ln\left(1+V_i^{\dagger}\left[U_0+y(U_1-U_0)
\right]V_i\right)\right|_{1p}
\nonumber\\
&&-V_i^{\dagger}{\mathrm{diag}}(\varepsilon_p/2,\varepsilon_p/2,0,
\varepsilon_p/2,\varepsilon_p/2)V_i,
\label{HEFF}
\end{eqnarray}
where $V_i={\mathrm{col}}( p^{\vphantom{\dagger}}_{i-\hat{x}/2,\uparrow},
p^{\vphantom{\dagger}}_{i+\hat{y}/2,\uparrow},
d^{\vphantom{\dagger}}_{i\uparrow},
p^{\vphantom{\dagger}}_{i-\hat{y}/2,\uparrow},
p^{\vphantom{\dagger}}_{i+\hat{x}/2,\uparrow})$ is the column of Fock-space
operators, referring to atoms centered on the $i$-th site, with $
d^{\vphantom{\dagger}}_{i\sigma}$ the copper operator. Explicitly,
$U_0={\mathrm{diag}}( e^{-\beta\varepsilon_p}-1, e^{-\beta\varepsilon_p}-1,
e^{-\beta U}-1, e^{-\beta\varepsilon_p}-1, e^{-\beta\varepsilon_p}-1) $, and
\begin{equation}
U_1-U_0=\left(
\begin{array}{ccccc}
 \tau_2 & -\tau_2 & \tau_1 & \tau_2 & -\tau_2 \\ -\tau_2 & \tau_2 & -\tau_1 &
 -\tau_2 & \tau_2 \\ \tau_1 & -\tau_1 & \tau_0 & \tau_1 & -\tau_1 \\ \tau_2 &
 -\tau_2 & \tau_1 & \tau_2 & -\tau_2 \\ -\tau_2 & \tau_2 & -\tau_1 & -\tau_2 &
 \tau_2 \\
\end{array}\right),
\end{equation}
with
\begin{subequations}
\label{tau}
\begin{eqnarray}
\tau_0&=&w\cos^2\frac{\varphi}{2}+\frac{p}{ w}\sin^2\frac{\varphi}{2}-c,\\
4\tau_1&=&-w\sin\varphi+\frac{p}{w}\sin\varphi,\\
4\tau_2&=&w\sin^2\frac{\varphi}{2}+\frac{p}{w}\cos^2\frac{\varphi}{2}-p,
\end{eqnarray}
\end{subequations}
where the following shorthand has been used:
\begin{subequations}
\label{sincos}
\begin{equation}
w=e^{\beta\widetilde{\Delta}_{pd}},\;p=e^{-\beta\varepsilon_p},\;c=e^{-\beta
U},
\end{equation}
and
\begin{eqnarray}
\sin\varphi&=&\frac{2t}
{\sqrt{\left(\frac{\varepsilon_p}{2}\right)^2+4t^2}},
\\
\cos\varphi&=&\frac{\varepsilon_p/2}
{\sqrt{\left(\frac{\varepsilon_p}{2}\right)^2+4t^2}},
\end{eqnarray}
while the exponent in $w$ is
\begin{equation}
\widetilde{\Delta}_{pd}=\sqrt{\left(\frac{\varepsilon_p}{2}\right)^2+4t^2}-
\frac{\varepsilon_p}{2}.
\label{deltapd}
\end{equation}
\end{subequations}

Coming back to $F_{\mathrm{eff}}$, the one-particle part of the logarithm may
be written explicitly, giving
\begin{subequations}
\label{Heff}
\begin{equation}
F_{\mathrm{eff}}=\sum_{i=1}^LV_i^{\dagger} TV_i,
\end{equation}
where
\begin{equation}
T=\left(
\begin{array}{ccccc}
\frac{\varepsilon_p}{2}+t_{pp} & -t_{pp} & t_{pd} & t_{pp} & -t_{pp} \\
-t_{pp}
 &\frac{\varepsilon_p}{2}+t_{pp} & -t_{pd} & -t_{pp} & t_{pp} \\ t_{pd} &
 -t_{pd} & \Delta\varepsilon_{d} & t_{pd} & -t_{pd} \\ t_{pp} & -t_{pp} &
 t_{pd} &\frac{\varepsilon_p}{2}+t_{pp} & -t_{pp} \\ -t_{pp} & t_{pp} &
 -t_{pd}
 & -t_{pp} &\frac{\varepsilon_p}{2}+t_{pp} \\
\end{array}\right).
\end{equation}
\end{subequations}
Before the expressions for the matrix elements of $T$ are given, note that
(\ref{Heff}) is like a simple tight-binding Hamiltonian, which may be
diagonalized by a Fourier transform. The spectrum consists of three bands, one
of which is nonbonding:
\begin{subequations}
\label{spec}
\begin{eqnarray}
\varepsilon_\pm(\gamma_k^2)&=& \frac{\varepsilon_p+\Delta\varepsilon_{d}}{2}
+2t_{pp}\gamma_k^2
\nonumber\\ &&\pm\;\sqrt{\left(\frac{\varepsilon_p-\Delta\varepsilon_{d}}{2}
+2t_{pp}\gamma_k^2\right)^2 +4t_{pd}^2\gamma_k^2},
\\
\varepsilon_0&=&\varepsilon_p,
\end{eqnarray}
\end{subequations}
with $\gamma_k^2=\sin^2k_x/2+\sin^2k_y/2$. The dispersion (\ref{spec}) is the
effective spectrum of up-spins in the presence of static down-spins. It was
denoted, schematically, as $\varepsilon_\alpha(k,\nu)$ in the definition
(\ref{zmunu}) of $Z(\mu,\nu)$.

In order to specify the model (\ref{model}) in full, it only remains to write
down the matrix elements of $T$ in (\ref{Heff}). Define $\zeta_\pm$ by
\begin{eqnarray}
e^{-\beta\zeta_\pm}&\equiv& \frac{p+c}{2}
+y\left(2\tau_2+\frac{\tau_0}{2}\right)
\nonumber\\ &&\pm\;\sqrt{\left[\frac{p-c}{ 2}
+y\left(2\tau_2-\frac{\tau_0}{2}\right)
\right]^2 +4y^2\tau_1^2},
\label{zeta}
\end{eqnarray}
where $y=1/(1+e^{\beta\nu})$, and the rest of the notation is as in
(\ref{tau},\ref{sincos}), while $\nu$ is, as always, the chemical potential of
the static down-spins. Then
\begin{subequations}
\label{hops}
\begin{eqnarray}
t_{pd}&=&\frac{\zeta_+-\zeta_-}{4}\sin\psi,
\label{tpd}
\\ 4t_{pp}&=&\frac{\zeta_++\zeta_-}{2}+\frac{\zeta_+-\zeta_-}{2}\cos\psi
-\varepsilon_p,
\\
\Delta\varepsilon_d&=&\frac{\zeta_++\zeta_-}{2}
-\frac{\zeta_+-\zeta_-}{2}\cos\psi,
\end{eqnarray}
\end{subequations}
where
\begin{eqnarray*}
\sin\psi&=&\frac{2y\tau_1}
{\sqrt{\left[\frac{p-c}{2}+y\left(2\tau_2-\frac{\tau_0}{2}\right)
\right]^2 +4y^2\tau_1^2}},
\\
\cos\psi&=&\frac{\left[\frac{p-c}{2}+y\left(2\tau_2-\frac{\tau_0}{2}\right)
\right]}
{\sqrt{\left[\frac{p-c}{2}+y\left(2\tau_2-\frac{\tau_0}{2}\right)
\right]^2 +4y^2\tau_1^2}}.
\end{eqnarray*}
This completes the derivation of the effective spectrum (\ref{spec}) of the
model (\ref{model}).

\subsubsection{Physical considerations}\label{phys}

Let us first summarize the physical truncations made in the preceding
derivation. Two distinct meanings of `first non-trivial term' were employed.
One is to keep to the one-particle part of the Fock space. This is guaranteed
by the Fock-space formalism itself, or more technically by the fact that one
has passed from operators to numbers only in the last step of the calculation,
when diagonalizing the operator~(\ref{Heff}). The other is to keep to first
order in the expansion in $y$. This is a little more tricky, because the final
spectrum~(\ref{spec}) of the free energy is non-linear in $y$. Here
`first-order' actually means first-order at the level of the evolution
operator. This is especially evident in the forms~(\ref{heff})
and~(\ref{HEFF}). Operationally, the rule is to keep to first order in $y$
under the logarithm, but the one-particle part of the resulting logarithmic
operator must be extracted exactly, \emph{i.e.} without further truncation in
$y$.

The above one-particle effective free energy is formally the first (one-site)
non-trivial term in an expansion around the atomic limit. In terms of the
expansion parameter $y=1/(1+e^{\beta\nu})$, where $\nu$ is the chemical
potential of the down-spins, this is an expansion around $y=0$, when the
copper sites are completely occupied by down-spins. It will now be shown that
the result is correct also for $y=1$, which means no down-spins at all
($\nu\to-\infty$). Then the exact non-interacting band is recovered. This is a
very useful check that the two truncation prescriptions discussed in the
previous paragraph are mutually consistent. In particular, it shows that
keeping to first order in $y$ does not violate the Pauli principle. This is a
benefit of working in Fock space --- in similar expansions around the
classical limit, the Pauli correlations are only taken into account in an
inclusion-exclusion procedure~\cite{Zhi91}, which remains approximate to all
finite orders in the expansion parameter.

Taking $\nu\to-\infty$ in Eq.~(\ref{zeta}), one gets
\begin{eqnarray}
e^{-\beta\zeta_+}\to w,&&e^{-\beta\zeta_-}\to p/w,\nonumber\\
\cos\psi\to -\cos\varphi,&&\sin\psi\to -\sin\varphi.
\label{limy1}
\end{eqnarray}
Putting all this together, the hopping amplitudes (\ref{hops}) become $$
t_{pd}\to t,\;t_{pp}\to 0,\;\Delta\varepsilon_d\to 0, $$ which gives the
dispersion for the free limit of the Hamiltonian (\ref{rt}). This confirms
that the one-particle part of the original problem has been treated correctly.
It is also easy to verify that putting $y=0$ gives the same result as $t=0$ in
the original Hamiltonian, but this is of course expected.

It will now be shown that the one-particle part of the expansion in $y$ is
controlled, for the concentrations considered in this article. The higher
order terms in $y$ contain one-particle contributions coming from the
diagonalization of larger `molecules,' such as Cu$_2$O$_7$ (say), which
appears when two `hopping crosses' share an oxygen. This has the typical
structure of a connected-cluster expansion. The general term is $y^pU_L(p)$.
For down-spin concentrations beyond half-filling, the saddle-point equations
give $\nu\approx\varepsilon_p$, \emph{i.e.} $y\sim e^{-\beta\varepsilon_p}=
e^{-\beta\Delta_{pd}}$ (remember we put $\varepsilon_d=0$). Note from
Eqs.~(\ref{HEFF}) and~(\ref{sincos}) that the magnitude of the term
multiplying $y$ is roughly $w=e^{+\beta\widetilde{\Delta}_{pd}}$, so the
overall estimate for the first term is $yU_L(1)\sim yw\sim
e^{-\beta(\Delta_{pd}-\widetilde{\Delta}_{pd})}<1$. Physically,
$\widetilde{\Delta}_{pd}$ is just the hybridization energy associated with the
diagonalization of a single CuO$_4$ `molecule.' A larger cluster will have a
larger hybridization gain, but it will also be suppressed by a larger power of
$y$. Since hybridization gains quickly saturate as a function of cluster size,
they cannot offset the addition of a whole bare $\Delta_{pd}$ with each new Cu
site added to a `molecule.' Hence the expansion is rapidly convergent in the
range of fillings considered below.

For down-spin fillings below the Mott transition, \emph{i.e.} corresponding to
the lower Hubbard band (LHB), $\nu\approx\varepsilon_d=0$, meaning $y$ is some
appreciable fraction of unity, not exponentially small. Then presumably
higher-order single-particle terms could play a role. This is consistent with
the fact that in the LHB, there is a significant contribution of the
interaction to the entropy, while the same is exponentially small beyond the
Mott transition --- the system goes from liquid to gas, absorbing latent
heat~\cite{Sunko96,Sunko00}. Nevertheless, the fact that the exact band result
is recovered in the limit $y=1$ suggests that the present calculation should
not be qualitatively incorrect even in the LHB, not of main interest here.

\section{Results}

\subsection{Small hopping overlap}

\begin{figure}[tb]
\center{\epsfig{file=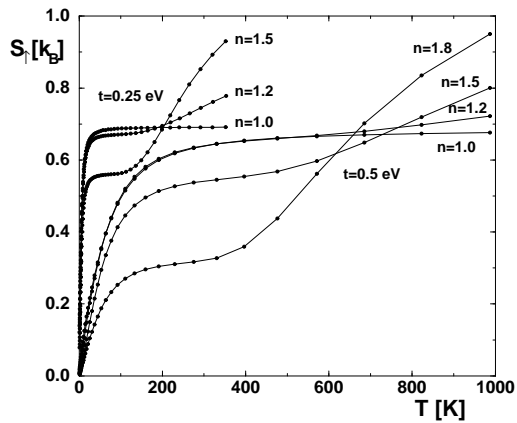,height=55mm}}
\caption{Entropy of mobile spins (\ref{entrop}), for $t=0.25$~eV (left three
curves), and $t=0.5$~eV (right four curves). Curves are marked by the
corresponding concentration. Saturation plateaus are clearly visible. Here
$\Delta_{pd}=3$~eV, $U=10$~eV, and $n=2n_\uparrow=2n_\downarrow$.}
\label{fig_enup_t}
\end{figure}

In the model, the entropy of mobile spins is formally given by the usual
single-particle expression
\begin{equation}
S_\up=-\int_{BZ}\left[f_\mu(\varepsilon_-)\ln(f_\mu(\varepsilon_-))
+(1-f_\mu(\varepsilon_-))\ln(1-f_\mu(\varepsilon_-))\right],
\label{entrop} \end{equation}
where $\mu$ is the chemical potential of the mobile spins, while the effective
bonding band $\varepsilon_-$, given by Eq.~(\ref{spec}), is a function of
$\nu$, the chemical potential of the static spins, and of the temperature.
Both $\mu$ and $\nu$ change with temperature, since the particle number is
fixed for the up- and down-spins separately. In Fig.~\ref{fig_enup_t}, the
entropy of mobile spins is shown as a function of temperature, for two values
of the bare hopping and several concentrations. The most striking feature of
the figure are broad plateaus, or inflections, indicating dips of width
$\sim$~100~K in the heat capacity $c_V$ of the system. The curve at $n=1.8$
even shows a `Kauzmann paradox'~\cite{Kauzmann48}, where extrapolation of the
high-energy part would give a negative entropy at zero temperature. Of course,
the entropy is strictly zero at $T=0$, since the free energy~(\ref{model}) is
microscopic by construction. The plateaus correspond to regions of low $c_V$,
associated with the difficulty of finding new states, which gives peaks in
$c_V$ (not shown) between $T=0$ and the plateau positions.

\begin{figure}[tb]
\center{\epsfig{file=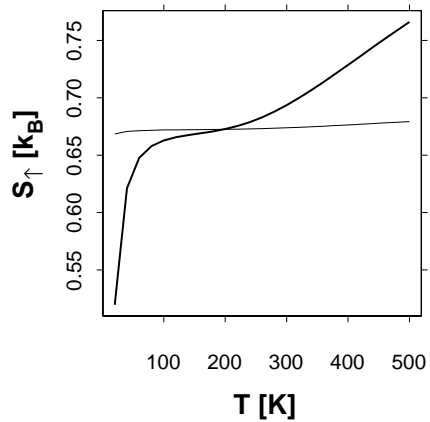,height=55mm}}
\caption{Comparison of $S_\uparrow(T)$ (thick line) and $S_{\uparrow,{\mathrm
Cu}}(T)$, Eq.~(\ref{SSCu}) (thin line), for $t=0.31$~eV, $n=1.2$, and other
parameters as in Fig.~\ref{fig_enup_t}.}
\label{fig_sscu_t}
\end{figure}

The plateaus are identified with static disorder in the obstructed copper
sites. The proof is as follows: call the concentration of up-spins on the
coppers $m_\uparrow$. If they were static, their contribution to the entropy
would be just 
\begin{equation}
S_{\uparrow,{\mathrm
Cu}}=-m_\uparrow\ln(m_\uparrow)-(1-m_\uparrow)\ln(1-m_\uparrow),
\label{SSCu}
\end{equation}
and beyond half-filling, each copper site would be singly occupied, either by
an up-spin or a down-spin. Thus $m_\uparrow=1-m_\downarrow$, in the static
case, where $m_\downarrow$ is the concentration of down-spins on the coppers.
Since $m_\downarrow$ is easily obtained from the calculation, one can use it
to construct $S_{\uparrow,{\mathrm Cu}}(T)$ from Eq.~(\ref{SSCu}), and compare
these values with the actual $S_\uparrow(T)$ from the same calculation, given
by Eq.~(\ref{entrop}). As shown in Fig.~\ref{fig_sscu_t}, the hypothetical
static entropy (\ref{SSCu}) coincides with the calculated entropy at the
inflection point, beyond which the latter begins to rise. To show this more
clearly, the hopping overlap is slightly increased in this figure, so the
plateau is not quite flat.

\begin{figure}[tb]
\center{\epsfig{file=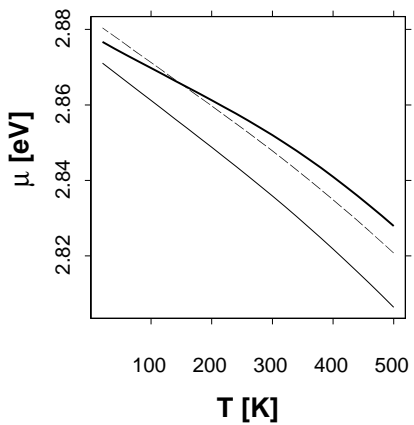,height=55mm}}
\caption{Chemical potential of mobile spins (thick line) corresponding to
$S_\uparrow(T)$ in Fig.~\ref{fig_sscu_t}. The bottom and top of the effective
band are given by thin and dashed lines, respectively.}
\label{fig_mu_t}
\end{figure}

This means that in the regime $t\ll\Delta_{pd}$, excitation of the system
proceeds in two steps. First the copper positional degrees of freedom are
fully saturated: their entropy cannot be greater than that of the correspondig
static disorder. Then the translational ones, involving the oxygens, are
activated, but only after a small ($\sim$~100~K) region is crossed, where
there are no new states. The two steps are better separated when the bare
hopping is smaller: note the difference between the curves for $t=0.25$~eV and
$t=0.5$~eV in Fig.~\ref{fig_enup_t}. 

Figure~\ref{fig_mu_t} shows how the model technically achieves this behavior.
As usual with single-particle effective models, collectivity enters through
the solution of the saddle-point equations, which fix the chemical potential
at a given concentration and temperature. The chemical potential traverses the
band fairly rapidly as the temperature increases, and leaves it roughly around
the temperature at which the plateau appears. As it passes through the region
where the 1$p$-DOS drops, there appears a lack of new states for the system.
Once in the (Maxwellian) tail of the Fermi distribution, the oxygen degrees of
freedom are further excited. In this sense, the system is again analogous to
$^3$He. However, there is no regime of classical kinetic motion in the model,
since the bands are built-in by assumption.

\subsection{Large hopping overlap}

\begin{figure}[tb]
\center{\epsfig{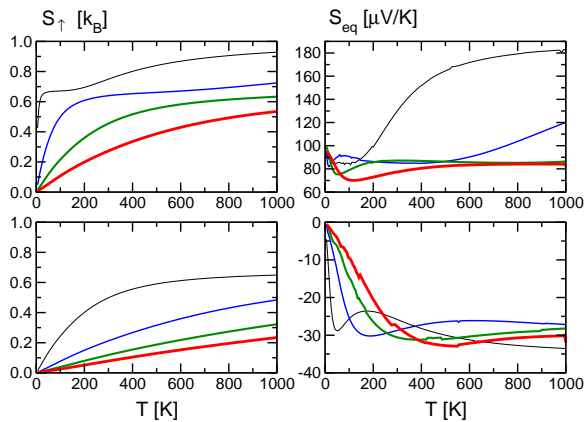}}
\caption{Left column: entropy of mobile spins (\ref{entrop}). Right column:
equilibrium thermopower (\ref{thermo}). Upper row: $n_\up=n_\dw=0.6$. Lower
row: $n_\up=0.6$, $n_\dw=0.001$. Lines, thin to thick: hopping
$t=0.25,0.5,0.75,1.0$ eV, respectively, with $\Delta_{pd}=3$ eV.}
\label{fig_kinet}
\end{figure}

Collective effects persist in the model even when the hopping overlap becomes
so large that the effective band appears to be normally metallic, with a
well-defined Fermi surface. They are no longer immediately evident at the
level of the entropy, but remain in the susceptibility. This is shown in
Figure~\ref{fig_kinet}. For comparison, the lower row shows the limit of free
fermions. The entropy for $n_\up=n_\dw=0.6$ (upper left panel) rapidly begins
to look `normal' as the hopping overlap increases (thicker curves).

The right column in the figure shows that this impression is misleading,
however. It gives the equilibrium susceptibility, corresponding to the
thermopower,
\begin{equation}
S_{eq}=-\frac{1}{e}\frac{\partial\mu}{\partial T}.
\label{thermo}
\end{equation}
In a simple rigid band, $S_{eq}$ should change sign from positive to negative
as the filling goes beyond one-half. This is because the chemical potential
goes down with the temperature, if the 1$p$-DOS goes up with the filling, and
vice versa.

For the free model (lower right panel), the equilibrium
thermopower~(\ref{thermo}) is indeed negative. Now add down-spins until the
model undergoes its Mott-Hubbard transition~\cite{Sunko00}. Then the up-spin
$S_{eq}$~(\ref{thermo}) becomes positive, even though the Fermi surface does
not change with respect to the free situation. It is physically clear what is
happening: the kinetic blocking effects are strongest around half-filling, so
going beyond it increases the available degrees of freedom, since the mobile
up-spins on the oxygens are not blocked. A similar outcome is expected from
the effect of antiferromagnetic (AF) correlations above the transition,
however there it would be traced to the appearance of a dip in the
single-particle spectrum. Namely, such a 1$p$-DOS has a minimum around
half-filling, so that beyond it, it rises again, just like at the bottom of
the band. Thus in the spin channel, an anomalous susceptibility is due to a
normal Fermi-liquid response. By contrast, in the charge channel and beyond
the Mott-Hubbard transition, the response cannot similarly be inferred from
the single-particle spectrum. Even though the model~(\ref{model}) is formally
single-particle and in the metallic regime, it is not a normal Fermi liquid.

\subsection{Heat capacity}

\begin{figure}[tb]
\center{\epsfig{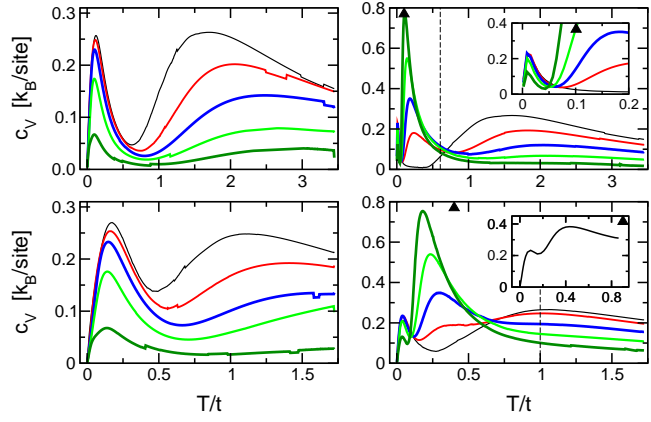}}
\caption{Heat capacities of the random-tiling model. Left (right) column: 
below (above) half-filling. Thin to thick: left,
$n=2n_\uparrow=2n_\downarrow=0.9,0.7,0.5,0.3,0.1$, right,
$n=2n_\uparrow=2n_\downarrow=1.0,1.2,1.4,1.6,1.8$, respectively. Upper row:
$t=0.5$~eV, $\Delta_{pd}=3$~eV. Lower row: $t=1$~eV, $\Delta_{pd}=3$~eV.
Vertical dashed line: renormalized charge-transfer scale
$\widetilde{\Delta}_{pd}$, Eq.(\ref{deltapd}). Insets: upper, detail of the
narrow peak; lower, $n=1.4$, $t=2$~eV, $\Delta_{pd}=3$~eV. Black triangles on
the axes: $W/t$, Eq.~\ref{bandw}.}
\label{fig_cv}
\end{figure}
The heat capacities of the random-tiling model are shown in Fig.~\ref{fig_cv}
(thinner lines are always closer to half-filling). In the LHB (left column),
there is a transfer of spectral intensity with doping, between a low- and
high-energy scale, both of which already exist at half filling. The latter is
still significantly lower than the bare charge-transfer scale $\Delta_{pd}$,
which takes the role of $U$ in the present context. This is an effect of
hybridization, as mentioned above: the larger the hopping overlap, the lower
this `high' scale becomes (note that the horizontal axes in the two rows are
different). Below the Mott-Hubbard transition, the equilibrium thermopower is
of the same sign as in the free case $n_\downarrow\to 0$
(Fig.~\ref{fig_kinet}), even for $2n_\uparrow=2n_\downarrow=0.9$.

Concentrations beyond half-filling cannot be studied in a one-band model. In
the present three-band setting, there appears another low-energy peak between 
the two peaks present at half-filling, which quickly draws strength from the
high-energy peak as the doping increases. The new low-energy scale induced by
doping is specific to the three-band model in the `in-gap' concentration
range. All the relevant scales can be extracted analytically in the limit
$T\to 0$, since in that limit the chemical potential of the down spins is
forced to one of two known values: $\nu\to\varepsilon_d=0$ in the lower Hubbard
band (LGB), and $\nu\to\varepsilon_p$ in the in-gap band (IGB), beyond the
Mott-Hubbard transition. The expressions (\ref{hops}) are easy to evaluate
in these limits. Below the transition, the width of the lower Hubbard
band is practically unchanged from the non-interacting value
\begin{equation}
W_0=\sqrt{\left(\frac{\Delta_{pd}}{2}\right)^2+8t^2}-\frac{\Delta_{pd}}{2}
\;\;\;\mathrm{(LHB)},
\label{bandw0}
\end{equation}
where $\Delta_{pd}=\varepsilon_p-\varepsilon_d$. Here the only effect of
strong correlations on the band is a small but finite shift of the copper
level downwards,
\begin{equation}
\lim_{T\to 0}\Delta\varepsilon_d=\Delta_{pd}\sin^2\frac{\varphi}{2}
-\widetilde{\Delta}_{pd}\cos^2\frac{\varphi}{2}<0\;\;\;\mathrm{(LHB)}.
\end{equation}
The IGB is, on the other hand, strongly affected by correlations. The copper
level is shifted upwards, and the band is significantly narrowed. Its width in
the zero temperature limit is~\cite{Sunko00}
\begin{equation}
\lim_{T\to 0}W=\frac{\widetilde{\Delta}_{pd}}{
1+\Delta_{pd}/(2\widetilde{\Delta}_{pd})}\;\;\;\mathrm{(IGB)}.
\label{bandw}
\end{equation}
The copper level shift is now large and positive:
\begin{equation}
\lim_{T\to 0}\Delta\varepsilon_d=\Delta_{pd}
-\widetilde{\Delta}_{pd}\cos^2\frac{\varphi}{2}>0\;\;\;\mathrm{(IGB)}.
\end{equation}
Instead of using this formula, it is much more convenient to note that the
distance from the bare oxygen position $\varepsilon_p$ to the \emph{middle}
of the IGB is just $\widetilde{\Delta}_{pd}$ in the same limit,
\emph{i.e.} the uppper edge of the IGB is given by any of the two formulas
\begin{equation}
\lim_{T\to 0}
\left(\varepsilon_d+\Delta\varepsilon_d\right)=
\lim_{T\to 0}\left(\varepsilon_p -\widetilde{\Delta}_{pd}+\frac{W}{2}\right)
\;\;\;\mathrm{(IGB)}.
\end{equation}

\begin{figure}[tb]
\center{\epsfig{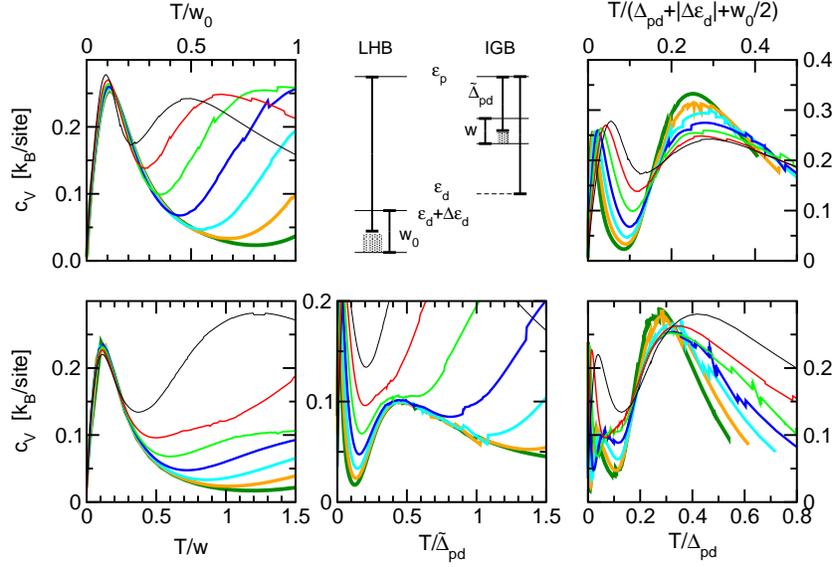}}
\caption{Scales of the peaks in heat capacities for $t=1$~eV. Top (bottom)
row:  n=0.9 (n=1.1). Lines, thin to thick: $\Delta_{pd}=2,3,4,5,6,7,8$~eV,
respectively. The drawing in the middle of the top row schematically shows the
scales as barred thick lines, left for the lower Hubbard band (LHB), right for
the in-gap band (IGB). The band edges are given by thin horizontal lines, and
the occupied levels indicated by shaded squares.}
\label{fig_scale}
\end{figure}
The relevance of these scales for the various peaks is analyzed in
Fig.~\ref{fig_scale}. Both below and above the Mott-Hubbard transition, the
lowest-energy peak is due to intraband transitions. The peak scales perfectly
with $W$ for the IGB, and with $W_0$ for the LHB. The high peak in the LHB is
due to interband processes, namely excitations into the empty, non-dispersive
non-bonding band at $\varepsilon_p$, which is a distance
$\Delta_{pd}+|\Delta\varepsilon_d|+W_0/2$ from the Fermi level when the
occupied band is nearly half-filled from below. The middle (new) peak in the
IGB is also due to interband transitions. As noted above, the middle of this
band is a distance $\widetilde{\Delta}_{pd}$ from the oxygen level, giving the
scale of the middle peak when the band is nearly half-filled from above.

Finally, the upper (third) peak in the IGB is again due to interband
transitions, but of a different kind. They are governed by the bare scale
$\Delta_{pd}$, which plays the role analogous to $U$. However, there is no
effective band in the model at $\varepsilon_d$ beyond the Mott-Hubbard
transition (the position $\varepsilon_d$ is denoted by a broken line for this
reason). The high peak is probing the `undressing' scale, at which the system
effectively reverts to the atomic limit. Note that the two thinnest lines,
corresponding to $\Delta_{pd}/t=2$ and 3, respectively, do not scale so well
by $\Delta_{pd}$ in the right column, and simultaneously have not yet
developed a third peak in the middle column. The separation of scales between
the middle and high peak should be regarded as a sign of a well-developed
charge-transfer regime $\Delta_{pd}>t$ and $n>1$, while for
$\Delta_{pd}\approx t$, and $n\approx 1$ from above, the two scales are mixed
by hopping fluctuations. The upper peak for these two thin lines has a higher
scale than $\Delta_{pd}$, because the copper level they see is still
effectively shifted by hopping correlations. This can be corroborated by an
amusing `quick and dirty' check (not shown), where one tries to scale the
upper peak in the IGB by $\Delta_{pd}+|\Delta\varepsilon_d|$, but using the
value $|\Delta\varepsilon_d|$ for the LHB. Even though this is formally
inconsistent, it turns out that the scaling in the lower right panel of
Fig.~\ref{fig_scale} is improved, since the level shift
$|\Delta\varepsilon_d|$ in the LHB is bigger for stronger hopping fluctuations
(smaller $\Delta_{pd}/t$).

The same two IGB regimes, hopping-fluctuation and charge-transfer, may be
discerned in the upper and lower panels in the right column of
Fig.~\ref{fig_cv}, drawn to absolute scale $t$. In the upper right panel, the
input $t/\Delta_{pd}\ll 1$, and the middle (new) peak is higher than $W$ and
lower than $\widetilde{\Delta}_{pd}$. Then $\widetilde{\Delta}_{pd}$ appears
as a crossover scale between the high-energy peak and the two lower-energy
ones. This is the charge-transfer regime, mentioned above.

The hopping-fluctuation regime prevails when the new peak is at low energy
with respect to both $W$ and $\widetilde{\Delta}_{pd}$. The lower right panel
in Fig.~\ref{fig_cv} shows the case $\widetilde{\Delta}_{pd}=t$, when some
strength still remains in the high-energy peak. The lowest peak has already
broadened, leaving no trace of the Brinkman-Rice effect (the scale of the
lowest peak has grown to about 1000~K, which is the whole range shown in
Fig.~\ref{fig_kinet}). The inset shows the hopping-fluctuation regime fully
developed,  where the upper peak has completely disappeared, so there are
again only two peaks like in the LHB (left column). Both are low-energy
phenomena ($\widetilde{\Delta}_{pd}/t\approx 1.39$ for the inset). For all of
the right column in Fig.~\ref{fig_cv}, the equilibrium thermopower stays of
the `wrong' sign, like in the upper row of Fig.~\ref{fig_kinet}, no matter how
large $t$ is. The fact that the metal is both strange and has an ordinary
Fermi surface is consistent with the fact that the unusual two-peaked feature
in the heat capacity is observed on a scale smaller than the band-width.

\subsection{Comparison with other work}

\begin{figure}[tb]
\center{\epsfig{file=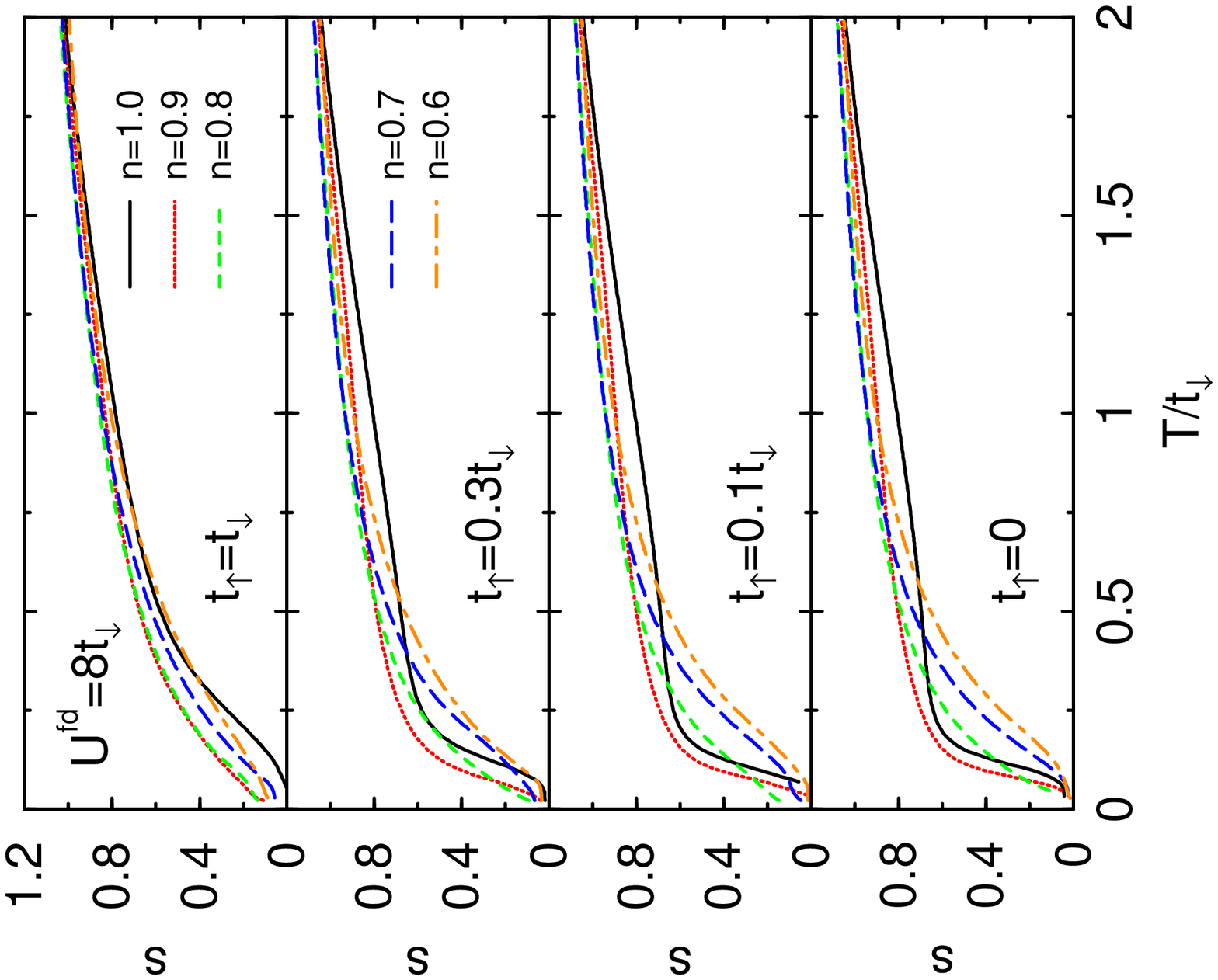,height=55mm,angle=-90,origin=c}
\epsfig{file=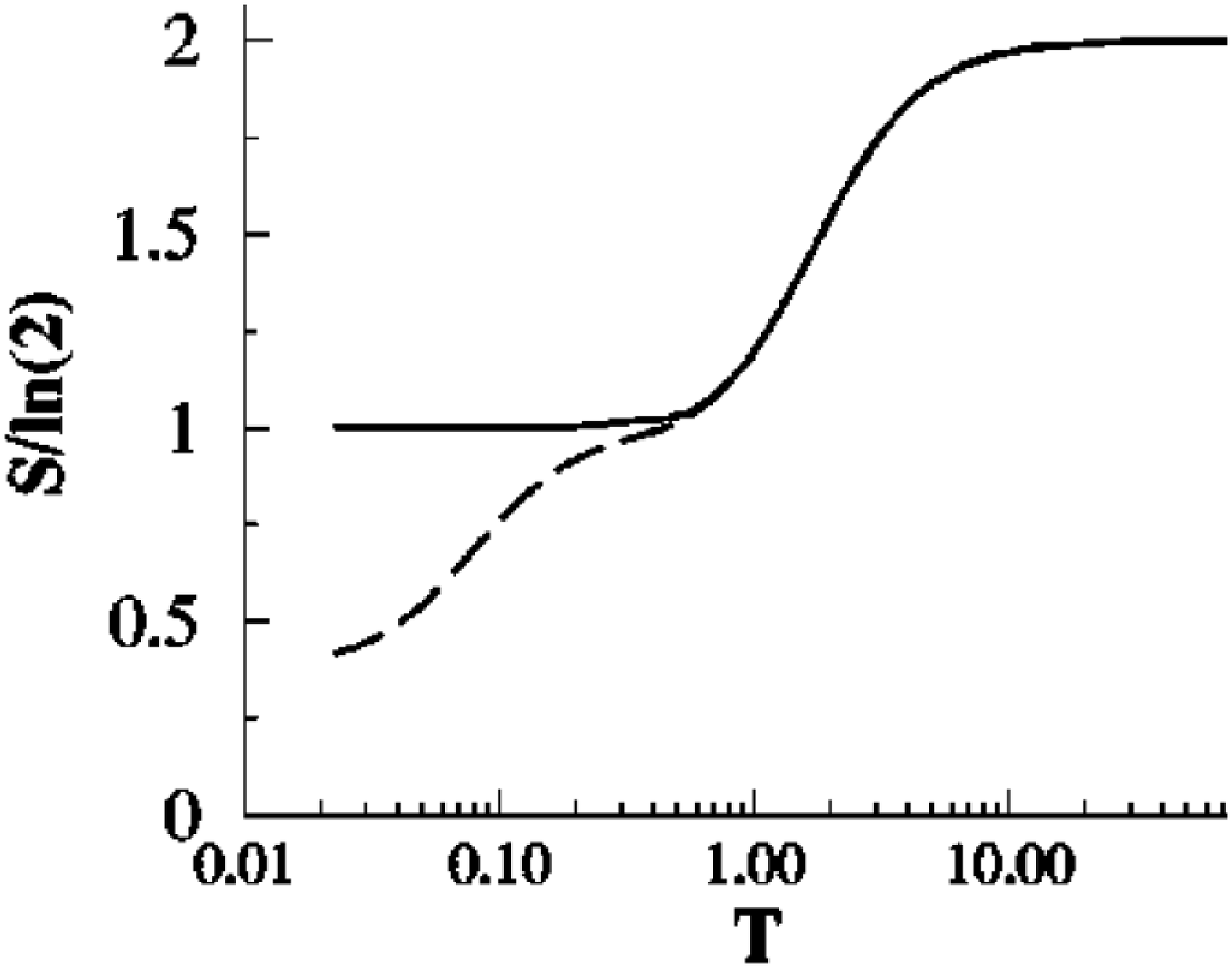,height=55mm}}
\caption{Left: numerical results with the Lanczos
algorithm~\cite{ElShawish03}. Right: the dynamical cluster
approximation~\cite{Hettler00}. For a detailed description, see the text.}
\label{fig_comp}
\end{figure}
To compare the present approach meaningfully with others, some general remarks
are in order. The approach described here works at present only for the
equilibrium properties, a limitation not shared by some other available
methods. It is in the thermodynamic limit, like other theoretical schemes, and
unlike purely numerical ones. It is unique by being fully analytic. The
spectrum~(\ref{spec}) is given in closed form, so the only numerical part of
the work is to fix the chemical potentials $\mu$ and $\nu$ by
Eqs.~(\ref{eqmunu}), like in a non-interacting band problem.

One should also have some idea, what such a comparison hopes to achieve. The
present paper regards models of Falicov-Kimball type as less interesting for
their own sake, than as stand-ins for problems one would really like to treat,
the dynamical Mott problem in particular. In this context there is no
\emph{a priori} advantage to one model over another, and it is preferable to
concentrate on their similarities. Differences are significant only if they
can be settled by a reliable external criterion, such as a known exact result.
Hence the purpose of this section is to identify some `generic' behavior of
this kind of models, see how the present work fits in, and hopefully
understand better the relationship to the Mott type of problem. The papers
used for comparison here are selected as representative of recent work,
without implication as to merit or priority relative to others.

El Shawish \emph{et al.}~\cite{ElShawish03} have studied a two-band spinless
Falicov-Kimball model, which may be mapped onto a Hubbard model in which the
two kinds of spins, up and down, hop with different overlaps, $t_\uparrow\neq
t_\downarrow$. The method was purely numerical, Lanczos algorithm on a lattice
of $4\times 4$ sites. The limit $t_\uparrow=0$ corresponds to the original
model~\cite{Falicov69}, and their result for this case is shown in the bottom
panel on the left in Fig.~\ref{fig_comp}. The full curve for half-filling is
directly comparable with the $n=1.0$ curves in Fig.~\ref{fig_enup_t}.
Obviously the general trends are the same, a plateau at $S=\ln 2$, dropping
steeply to $S=0$ at $T=0$, and rising very slowly at high temperatures, due to
incoherent hopping. However, in Fig.~\ref{fig_comp} the entropy curves show a
gap in the spectrum, while in Fig.~\ref{fig_enup_t} there is only a tendency
for the effective mass to diverge, or band-width to collapse, like in the
Brinkmann-Rice transition, broadened here to a crossover. Recall that the
first peak in the heat capacity is scaled by the band-width $W$, which is very
small when $t\ll\Delta_{pd}$, meaning the entropy plateau is pulled in to very
low temperature instead of being gapped. The gap is missing because only the
one-particle part in Fock space is kept; one expects that the two-particle
terms would indeed produce a gap near half-filling, once the band-width in the
one-particle sector became small enough. (It is interesting that the down-spin
entropy in the present model goes to zero at the Mott
tranisition~\cite{Sunko96}, indicating an ordered pattern, even in the absence
of two-particle terms.)

The other panels in the left of Fig.~\ref{fig_comp} show the evolution in
$t_\uparrow/t_\downarrow$, up to the full Hubbard model
$t_\uparrow=t_\downarrow$ in the top panel. The evolution from small
$t/\Delta_{pd}$ to large $t/\Delta_{pd}$ in the upper left panel of
Fig.~\ref{fig_kinet} is remarkably similar. This could have been expected,
because the two spin orientations correspond to the $f$ and $d$ bands in the
original two-band spinless Falicov-Kimball model. Hence an equality between
the $f$ and $d$ bandwidths is naturally similar to a small (in units of $t$)
splitting of the copper and oxygen energies in the present work. The
similarity nevertheless raises some issues in relation to the Mott problem,
which will be touched upon in the Discussion below.

Hettler \emph{et al.}~\cite{Hettler00} have studied the two-dimensional
one-band Falicov-Kimball model by the dynamical cluster approximation (DCA).
This is an extension of dynamical mean-field theory (DMFT~\cite{Georges96}) to
clusters of more than one site, embedded in the same, spatially unresolved
host medium. In infinite dimensions, the simplest, one-site DCA reduces by
construction to the DMFT~\cite{Georges96,Hettler00}. In finite dimensions, it
reduces to the latter's finite-dimensional analogue, the dynamical mean-field
approximation (DMFA). The DCA results~\cite{Hettler00} for the entropy are
shown in the right panel of Fig.~\ref{fig_comp}, for a one-site cluster (DMFA,
full line), and four-site cluster (broken line). The DMFA retains the full
$\ln 2$ per site entropy in the ground state. The four-site DCA does better,
but still appears to saturate at about one-half that value near $T=0$. Hence
the DCA agrees with the present work, and with the numerical Lanczos results
above, in the high-temperature regime, but does not cross over correctly to
low temperature.

\begin{figure}[tb]
\setlength{\unitlength}{1mm}
\begin{picture}(100,55)
\put(0,5){\epsfig{file=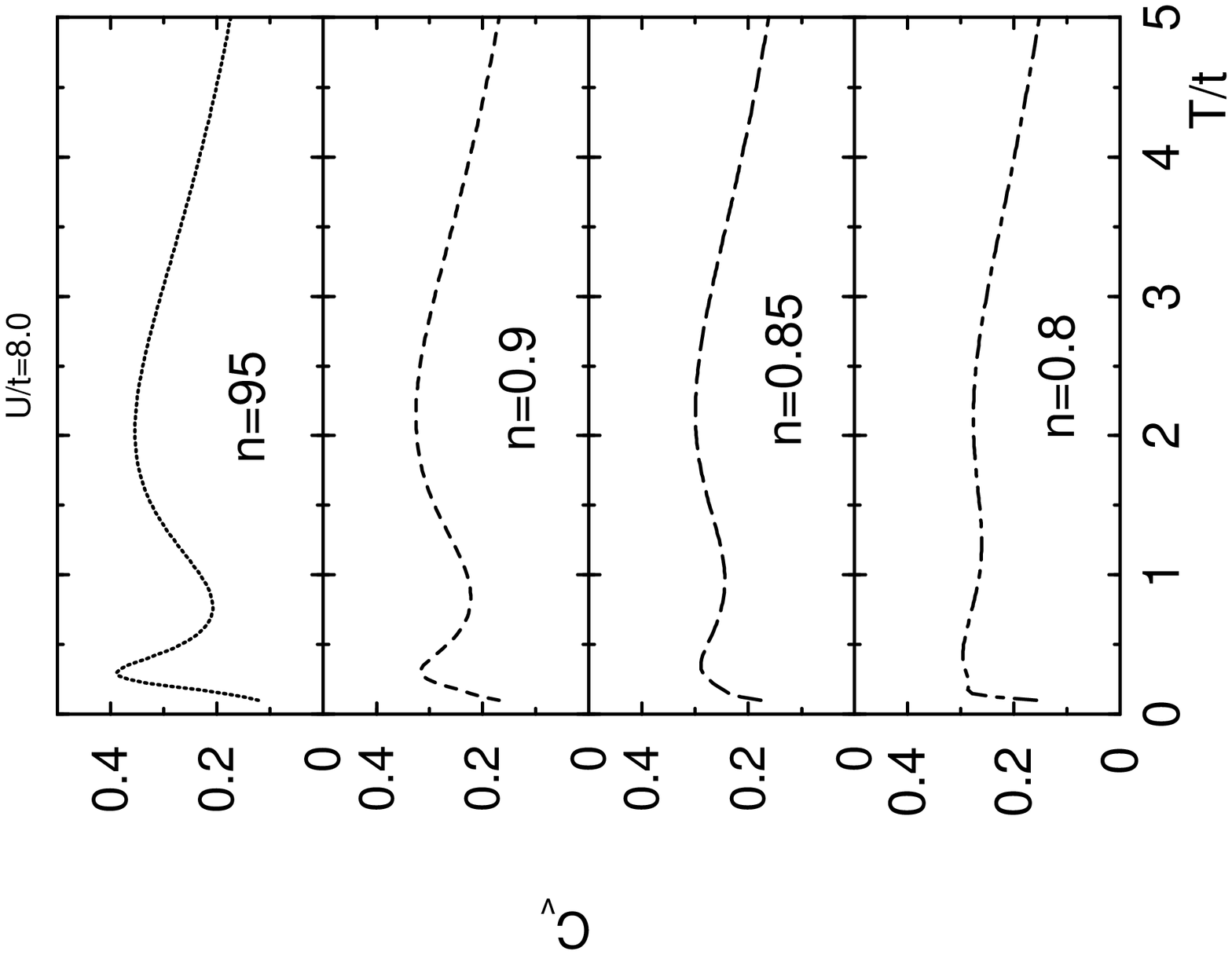,height=45mm,angle=-90,origin=c}}
\put(50,0){\epsfig{file=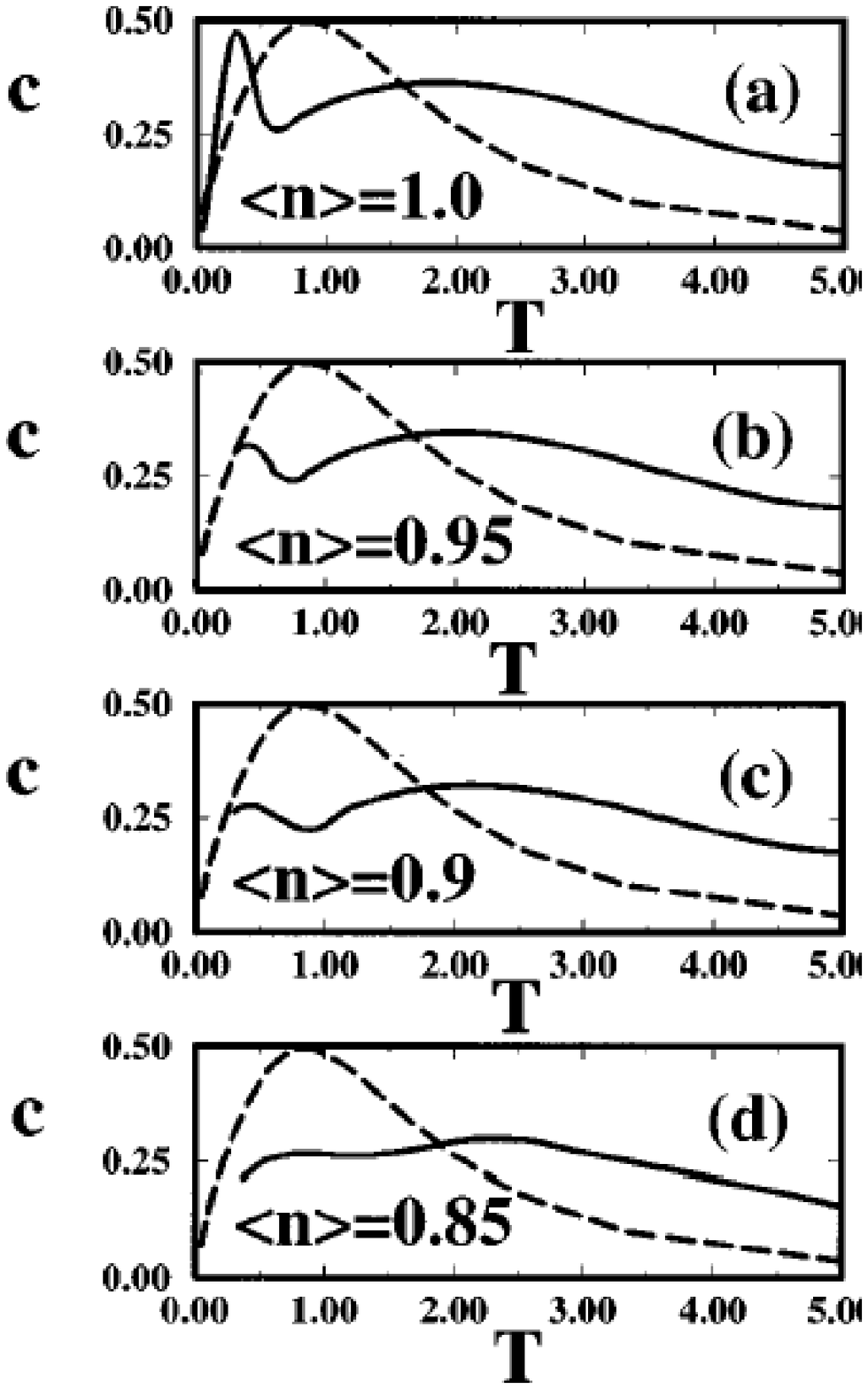,height=55mm}}
\end{picture}
\caption{Heat capacities of the Hubbard model, from Lanczos~\cite{Bonca03}
(left) and quantum Monte Carlo~\cite{Duffy97} (right) calculations, discussed
in the text. Note that positions of panels with the same concentrations are
shifted.}
\label{fig_cv_hubb}
\end{figure}
The heat capacities for the IGB in the right column of Fig.~\ref{fig_cv} cross
at two fairly well defined points. Vollhardt~\cite{Vollhardt97} has noted this
to be a quite general feature of strongly correlated systems, for families of
heat capacity curves parametrized by pressure or concentration in experiment,
or the on-site repulsion $U$ in the one-band Hubbard model. The crossing of
concentration-parametrized families of curves in the IGB, and their failure to
cross in the LHB, fits well with previous observations based on
Fig.~\ref{fig_kinet}, that it is the IGB, rather than the LHB, which is
singularly affected by strong correlations.

A two-peak structure in $c_V$ is ubiquitously found in numerical studies of
Hubbard-like systems~\cite{ElShawish03,Bonca03,Hettler00,Duffy97}. In
Fig.~\ref{fig_cv_hubb} on the left are the results of a Lanczos
calculation~\cite{Bonca03} for the one-band Hubbard model on a $4\times 4$
lattice. The right-hand panel is from a quantum Monte Carlo (QMC) calculation
on a $6\times 6$ lattice for the same model. They do not cross at any one
point, and have two peaks, both of which are generically features of the LHB
in the present model. This is not surprising, since concentrations below
half-filling are the only ones relevant in the one-band model, because of
particle-hole symmetry.

It has been noted~\cite{Bonca03} that the two calculations in
Fig.~\ref{fig_cv_hubb} differ significantly in their outcomes. First, in the
QMC results the first peak loses strength much faster with decreasing doping.
Second, in the QMC results the peak also moves, while in the Lanczos
calculations it stays much in the same place. From the RT model's point of
view, the first peak is expected to scale with the non-interacting bandwidth
$W_0$, so its position should not depend on concentration, except insofar as
due to the changing position of the Fermi level. This is visible in the left
column of Fig.~\ref{fig_cv}. The first peak there barely moves, but it moves a
little more in the bottom left panel, where the non-interacting band is wider,
so the Fermi level shifts more rapidly with concentration. Also, the narrower
the LHB, the sharper the first peak, and the more persistent as the
concentration decreases. This suggests a physical interpretation of the
differences in Fig.~\ref{fig_cv_hubb}. The QMC heat capacities imply that the
effective band seen by one-band Hubbard electrons below half-filling rapidly
widens as the concentration decreases. The Lanczos calculation implies a
situation similar to the present model in the LHB: band-width independent of
concentration, and the effective band somewhat narrower than the QMC
prediction.

To summarize, below half-filling the present analytical method gives similar
qualitative behavior of thermodynamic quantities as other, numerically much
more involved treatments of one-band models. It is also quantitatively correct
in the known exact limits, of high and low temperature, and high and low
doping. Where it can be checked quantitatively in the intermediate regime,
namely in Fig.~\ref{fig_sscu_t}, it is correct as well. It misses the gap in
the up-spin spectrum around half-filling, replacing it with a BR bandwidth
collapse, presumably because it retains only one-particle terms in the
Fock-space free energy operator. The doping dependence above half-filling is
qualitatively different than expected from the one-band analogy. At fixed
Hamiltonian parameters, there is a transfer of spectral weight to a new scale,
which emerges only with doping. The new scale is connected with interband
excitations across the renormalized charge-transfer gap. The present
calculation also parallels Lanczos calculations on the Hubbard model, both in
the shape and behavior of the peaks in the heat capacity below half-filling.

\section{Discussion}

The main modelling interest of the present article was to determine when the
one-particle construction, adopted to calculate the free energy, indeed
corresponds to a Fermi liquid, and when it does not. For that purpose, a
minimal negative criterion to observe a non-Fermi liquid at a finite
temperature was adopted. A system is not a Fermi liquid if the
susceptibilities do not follow from the 1$p$-DOS at that temperature. Put
conversely, a necessary condition for a system to be a Fermi liquid is that
the 1$p$-DOS responds `rigidly' to an infinitesimal change in temperature.

It has been found previously~\cite{Sunko00} that the Mott transition in the RT
model is first order, from a liquid of light particles, in the lower Hubbard
band, to a gas of heavy ones, in the renormalized in-gap band. (While the
down-spin entropy drops to zero at the transition, indicating AF order, the
chemical potential is a monotonous function of the filling, \emph{i.e.} there
is no phase separation.) In the present work, it is shown that while the
liquid is a Fermi liquid, the gas is not. The absence of residual interactions
(which makes it a gas) has been bought at the price of an anomalous variation
of the effective mass with the temperature, such that the thermopower
susceptibility (\ref{thermo}) cannot be predicted from the 1$p$-DOS at any
given temperature. This remains so even if the hopping overlap is large,
because the model is always in the strong-coupling limit: changing the hopping
does not affect the geometric obstructions, as obvious from Eq.~(\ref{rt}).
The collectivity involved is very simple. The fermions need to agree which are
the best positions for the annealed impurities, in precise analogy to $^3$He
atoms pushing against each other.

At low hopping overlap, this geometric effect so dominates the mean properties
of the gas, that it is unable to find new states of motion in a range of
temperatures, giving rise to plateaus in the entropy. (Extrapolating
experimental trends~\cite{Seiler86}, one could imagine similar plateaus
appearing in $^3$He around 0.5~K, if the pressure were sufficiently
increased.) These are remarkably similar to Kauzmann plateaus in vitreous
liquids, where they are due to the configurational rearrangements falling out
of equilibrium. Clearly the Kauzmann phenomenon~\cite{Kauzmann48} is a
non-specific sign that kinetic exploration of configuration space has been
obstructed, whatever the responsible mechanism. Here it has an interesting
connection with the Brinkmann-Rice (BR) bandwidth collapse~\cite{Brinkman70},
which is signalled in Fig.~\ref{fig_enup_t} by a steep rise of the entropy
from zero temperature at half-filling, indicating an infinite effective mass.
By the same token, the entropy plateau means that the effective mass is zero,
and it is obtained by continuous deformation of the BR entropy curves with
doping. It may seem surprising that obstruction of kinetic motion should be
associated with zero thermodynamic effective mass, but note that in the
absence of new states, there can be neither conduction nor dissipation. Zero
effective mass is just the equilibrium version of what is observed in the Hall
effect: the longitudinal Hall conductance and resistance simultaneously drop
to (almost) zero at the steps in the transverse resistance, precisely because
most of the carriers are localized~\cite{Klitzing86}.

The place of the present calculation among various cluster approaches is
simple to state. It is the ordinary Kubo cluster cumulant expansion of the
annealed free energy, with the single proviso that it is calculated in Fock
space. The transition from operators to numbers is made at the very end, after
the single-particle Fock term has been extracted. It is theoretically quite
revealing that the results presented here could be obtained by diagonalizing a
single CuO$_4$ molecule, \emph{i.e.} the lowest non-trivial term in the
underlying cluster expansion. Clearly, the main work was in fact done by the
algebraic and combinatorial machinery which respectively enforced both the
Pauli principle and the geometric constraint. Once these kinematical issues
were properly separated from the dynamical ones, it turned out the problem was
dominated by the former. This outcome is physically plausible by adiabatic
arguments: the low-energy physics of the Falicov-Kimball model is not
controlled by any elaborate dynamical correlations, such as one would expect
in the Mott case, where the scatterers and the scattered are equally light. It
is also consistent with the previous observation~\cite{Sunko00} that the
system is a gas beyond half-filling. In the lower Hubbard band, where it is a
liquid, higher-order clusters should affect the results, but that is not
expected to be qualitatively significant.

The comparison with other, numerically much heavier calculations in the
preceding section gives some grounds for reflection on the Mott problem. This
has been dominated for the past forty years by Gutzwiller's
intuitions~\cite{Gutzwiller65}, that electrons of one spin see those of the
other as a `smeared background' (Gutzwiller ansatz), and further `as if
occupying a band of width zero' (Gutzwiller approximation). This point of view
clearly implies there should be some similarity between Falicov-Kimball and
Hubbard-model low-energy behavior. Such a similarity appears to have been
found in one of the calculations cited above~\cite{ElShawish03}, in the smooth
evolution of the entropy as $t_\uparrow/t_\downarrow$ was raised from zero to
one. If one knew that Gutzwiller's proposals were correct, the very similar
evolution of the entropy, observed in the present calculation as
$t/\Delta_{pd}$ was increased, would be naturally expected. As things stand,
one should keep in mind the alternative resonating-valence-bond (RVB) scenario
of Anderson~\cite{Anderson87}, in which quantum coherence between the two spin
orientations is an essential ingredient of Mott's insulating ground state.
Thus the possibility remains open, that the Falicov-Kimball limit acts as a
kind of trap for all approximate treatments of the Hubbard model which
incidentally destroy this coherence, over time or space.

It is also of some interest that the heat capacities in the RT model are
qualitatively more similar to numerical Hubbard model results, than to
one-band Falicov-Kimball results. It is possible that the oxygen degree of
freedom plays a role analogous to the `other' spin in the one-band Hubbard
model.

To conclude, geometric obstruction in thermodynamic equilibrium has been
explored in an analytic single-particle quantum model of Falicov-Kimball type.
In the regime $t\ll\Delta_{pd}$, it gives rise to a plateau in the entropy of
mobile spins. For $t\sim\Delta_{pd}$, the mobile spins appear to be normally
metallic, only with renormalized parameters. Beyond half-filling, their
equilibrium thermopower susceptibility nevertheless reflects the microscopic
kinematic restrictions, behaving oppositely to what is expected from the
single-particle properties. For these fillings, correlation effects in the
model do not simplify in the limit of low temperature and large hopping
overlap.

\section{Acknowledgements}

Conversations with S.~Bari\v{s}i\'{c} and E.~Tuti\v{s}, and one with
T.M.~Rice, are gratefully acknowledged. Thanks are due to J.~Bon\v ca for
providing the left panels of Figs.~\ref{fig_comp} and~\ref{fig_cv_hubb}. This
work was supported by the Croatian Government under Project 0119256.

\end{document}